\documentclass[%
 twocolumn,reprint,
 amsmath,amssymb,
 aps,
]{revtex4-2}

\usepackage{graphicx}
\usepackage{dcolumn}
\usepackage{bm}
\usepackage{bbm}
\usepackage{amsmath}
\usepackage{amssymb}
\usepackage{float}

\begin{document}
\preprint{APS/123-QED}

\title{Information Synergy Maximizes the Growth Rate of Heterogeneous Groups}
\author{Jordan T. Kemp$^1$, Adam G. Kline$^1$, and Lu\'is M. A. Bettencourt$^{2,3}$}
 \affiliation{%
 $^1$Department of Physics, University of Chicago, Chicago, Illinois 60637, USA}%
\author{}%
\affiliation{%
 $^2$Department of Ecology and Evolution, University of Chicago, Chicago, Illinois 60637, USA}%
\affiliation{%
 $^3$Mansueto Institute for Urban Innovation, University of Chicago, Chicago, Illinois 60637, USA}%
\date{\today}

\begin{abstract}
Collective action and group formation are fundamental behaviors among both organisms cooperating to maximize their fitness, and people forming socioeconomic organizations. Researchers have extensively explored social interaction structures via game theory and homophilic linkages, such as kin selection and scalar stress, to understand emergent cooperation in complex systems. However, we still lack a general theory capable of predicting how agents benefit from heterogeneous preferences, joint information, or skill complementarities in statistical environments. Here, we derive general statistical dynamics for the origin of cooperation based on the management of resources and pooled information. Specifically, we show how groups that optimally combine complementary agent knowledge about resources in statistical environments maximize their growth rate. We show that these advantages are quantified by the information synergy embedded in the conditional probability of environmental states given agents' signals, such that groups with a greater diversity of signals maximize their collective information. It follows that, when constraints are placed on group formation, agents must intelligently select with whom they cooperate to maximize the synergy available to their own signal. Our results show how the general properties of information underlie the optimal collective formation and dynamics of groups of heterogeneous agents across social and biological phenomena.
\end{abstract}

\maketitle

\section{Introduction}

  Collective behavior is a general feature of biological and social systems.  It mediates the survival and evolution of populations under resource constraints, competition, or predation in natural systems ~\cite{pennisi2009origin} and the formation and persistence of social organizations in human societies \cite{stinchcombe2000social}. 
  Much past work has modeled collective dynamics using homogeneous interaction rules, common to all agents, that are often also phenomenological. While these models have produced diverse insights, they typically lack a theoretical foundation to explain how specific social behavior emerges among individual agents with heterogeneous information and behavior. Thus, there remain significant knowledge gaps in most realistic situations, where agents with distinct but potentially complementary traits act collectively to maximize their joint growth (fitness, wealth) in knowable, noisy environments. 

  Some examples help illustrate the present situation. Game theorists and ecologists have considered many different cooperative interaction schemes~\cite{sachs2004evolution} and explored evolutionary stable behavior~\cite{hammerstein1994game}, particularly on networks \cite{shakarian2012review,perc2013evolutionary,jackson2015games}, where optimal behavior is identifiable under given interaction rules. Elaborating these schemes by introducing higher order interactions has broadened our understanding of more complex social networks \cite{gomez2012evolution,battiston2021physics,alvarez2021evolutionary}, and their dynamical phase-stability under varying interaction strengths~\cite{ ferraz2021phase}. Researchers have also studied, both theoretically and in the laboratory, how memory of previous interactions  influences agents' preferences for future encounters  \cite{mccabe1996game,sachs2005experimental,goyal2005network,gracia2012heterogeneous}, the spread of social crises across distance \cite{lee2011impact}, and the formation and scaling properties of social collectives \cite{rand2011dynamic,castellano2000nonequilibrium}, such as cities \cite{bettencourt2013origins,schlapfer2014scaling}.

    In addition to interaction rules and associated payoffs, collective dynamics is predicated on maximum principles, which specify agents' preferences in view of a goal and thus render their behavior intelligent (optimal). For example, inclusive fitness theory, which assumes a reproductive benefit to cooperation because of shared genes~\cite{hamilton1963evolution,pepper2000relatedness} has been studied in mixing populations and over networks \cite{ohtsuki2006simple} where it predicts population benefits to cooperation through several forms of reciprocity~\cite{queller1985kinship}. More recently, researchers have studied resource pooling in models of growth as a means to minimize environmental uncertainty and associated loss of fitness among agents experiencing independent fluctuations with shared statistics~\cite{peters2022ergodicity,lightner2023need}. 
    Such approaches remain limited by the association between collective behavior and (genetic) homophily but they can help explain the existence of phase transitions in cooperation networks \cite{castellano2000nonequilibrium, ferraz2021phase}, and specify agents' plausible  behavioral patterns~\cite{goyal2005network}, even if doubts remain about inclusive fitness's predictive power \cite{nowak2017general}. 
    
    Generally, however, most current quantitative frameworks fail to address collective dynamics  when agents remain heterogeneous across skills, knowledge, and behavior~\cite{pepper2002mechanism,stinchcombe1990information}.
    Developing more general approaches to collective behavior that include adaptation and learning along with heterogeneity, is a crucial step towards understanding how agents self-organize in more complex and dynamical environments, where specialization and the division of labor and knowledge become key.

Adaptive behavior requires agents to acquire and process information over time \cite{dooley1997complex,frank2012natural} in response to their environments and to each other.
In any realistic situation, limited experience, specialization costs and physical limitations of effort, energy and time, all prevent agents from perfecting their knowledge of complex environments~\cite{miller1956magical,sweller1988cognitive}. A natural way to mitigate these individual limitations is to pool knowledge across agents leading to the formation of social organizations \cite{hume2003treatise}, and the division and coordination of labor in terms of their behavior~\cite{cooper2018division}. This is widely observed in human organizations, but also in animal social behavior starting with the division of labor by age and sex.

By working jointly to predict characteristics of their environment~\cite{stinchcombe1990information} and gather resources, groups of agents can maximize their collective fitness even when each individual has very limited knowledge. In a setting where there are resource returns to successful prediction and behavior, information of the state of a statistical environment determines the fitness of the population~\cite{kemp2023learning,bettencourt2019towards}, though there are questions about how such benefits  emerge quantitatively~\cite{bettencourt2009rules}. 
Here we formalize the calculation of these social benefits in terms of the properties of information and show how maximizing knowledge complementarities (synergy) maximizes the long-term growth rate of collectives. Specifically, we derive an expression for the additional payoff to cooperative behavior in terms of the joint information synergy about the agents' dynamical environment. 

These results lead us to introduce the principle of maximum synergy, which maps the maximization of collective resource growth rates into optimal social interaction structures. This work adds new dimensions to the study of collective  dynamics by connecting the structure of groups to that of information in complex environments mediated by agents' diverse subjective characteristics, such as their present knowledge and their life histories.

\section{Theory of Collective Growth}

We start by demonstrating how the benefits of collective action emerge from pooling information in synergistic situations. Synergy means the combination of behavior, knowledge, and skills that complement each other towards a goal. This concept is necessary for creating effective  organizations that embody complex information \cite{stinchcombe1990information}, but it is often not sufficiently formalized in common language, such as in discussions of innovation~\cite{fuller1982synergetics} or firm structure.

Here, we will refer to synergy as an explicit information theoretic quantity that measures the additional predictive power that a group has upon pooling its agent's information together, relative to the knowledge of each individual separately. This quantity has been introduced sometime ago in the context of studying circuits in information processing systems~\cite{schneidman2003synergy,bettencourt2007functional}, and has provided a framework for studying  higher-order neuron interactions in the brain \cite{varley2023multivariate}, and causality and information in complex systems \cite{mediano2022greater,varley2022emergence}. As we will show, synergy results formally from the conditional dependence between the probability of predictive signals distributed in a population and events in a shared environment.  
The gain in predictive power from agents pooling information as collectives allows them to obtain additional resources from a knowable environment beyond what agents alone can do, thus boosting their relative fitness or productivity.

It follows that collectives that seek to maximize their resources over long times must combine the information from their agents' individual models of the world in a way that accesses the most synergy. Groups that do not know \textit{a priori} how to realize their synergies must adjust their collective knowledge and interaction structure by observing outcomes of their environment in an iterative learning process.
After developing the general framework for group formation and collective growth across group sizes, we demonstrate a model environment that exhibits synergy using logic gates that take signals as inputs, and output probabilistic events.
We will demonstrate how synergy scales with the number of unique signals in a collective, and how specific  combinations of signals affect the average growth of resources for the group.

\subsection{Collective Growth in Synergistic Environments}

We consider a population of $N$ agents, each with initial resources $r_i$ that can be (re)invested into the set of outcomes of their environment to generate returns.
Each agents has access to a private signal, $s_j\in S_j$, which is used to predict the state of the environment and makes resource allocations  to possible outcomes $e\in E$. This signal may represent a number of different processes such as sensory input, or a lead retrieved from memory. 
With accurate parameterization of a model of the environment, $P(E|S_j)$, an agent's optimal investment strategy leads to a resource growth rate $\gamma=I(E;S_j)$ \cite{kemp2023learning}.
Agents with better models (and  better statistical estimations) experience higher average growth rates given by the
information rate of the agent's signal for the environment.

We now define our environment by a set of $l$ signals with unique statistics, $\boldsymbol S\equiv\{S_1,\dots S_l\}$ as $P(E|\boldsymbol S)$, with marginals of events $P(E)$ and signals $P(\boldsymbol S)$.
The joint information that $\boldsymbol S$ has on $E$ is at least equal to $S_j$, that is $I(\boldsymbol S;E)\geq I(S_j;E)$.
Generally, this inequality is strict if the conditional information $I(\boldsymbol{S}|E)>I(\boldsymbol S)$
~\cite{schneidman2003synergy,bettencourt2007functional}. 
We compute the total information by summing over the mutual information between the signals independently, subtracted by an interaction term across signals,
\begin{equation}\label{eqn1}
    I(E;\boldsymbol S)=\sum_jI(E;S_j)-R_P.
\end{equation}
\noindent The quantity, $R_P$, denoted the coefficient of redundancy, measures the strength of this conditional dependence across larger sets of signal (two, three, etc way). It is defined in App. \ref{infoagg}
\begin{equation}\label{generalRedundancy}
\begin{split}
    R_P
    =&\sum_{j>k=1}^lR(E;S_j;S_k)+\sum_{j>k>m=1}^lR(E;S_j;S_k;S_m)\\
    &+\dots+R(E;S_1;\dots;S_l).
    \end{split}
\end{equation}

\begin{figure*}
    \centering
    \includegraphics[width=.82\textwidth]{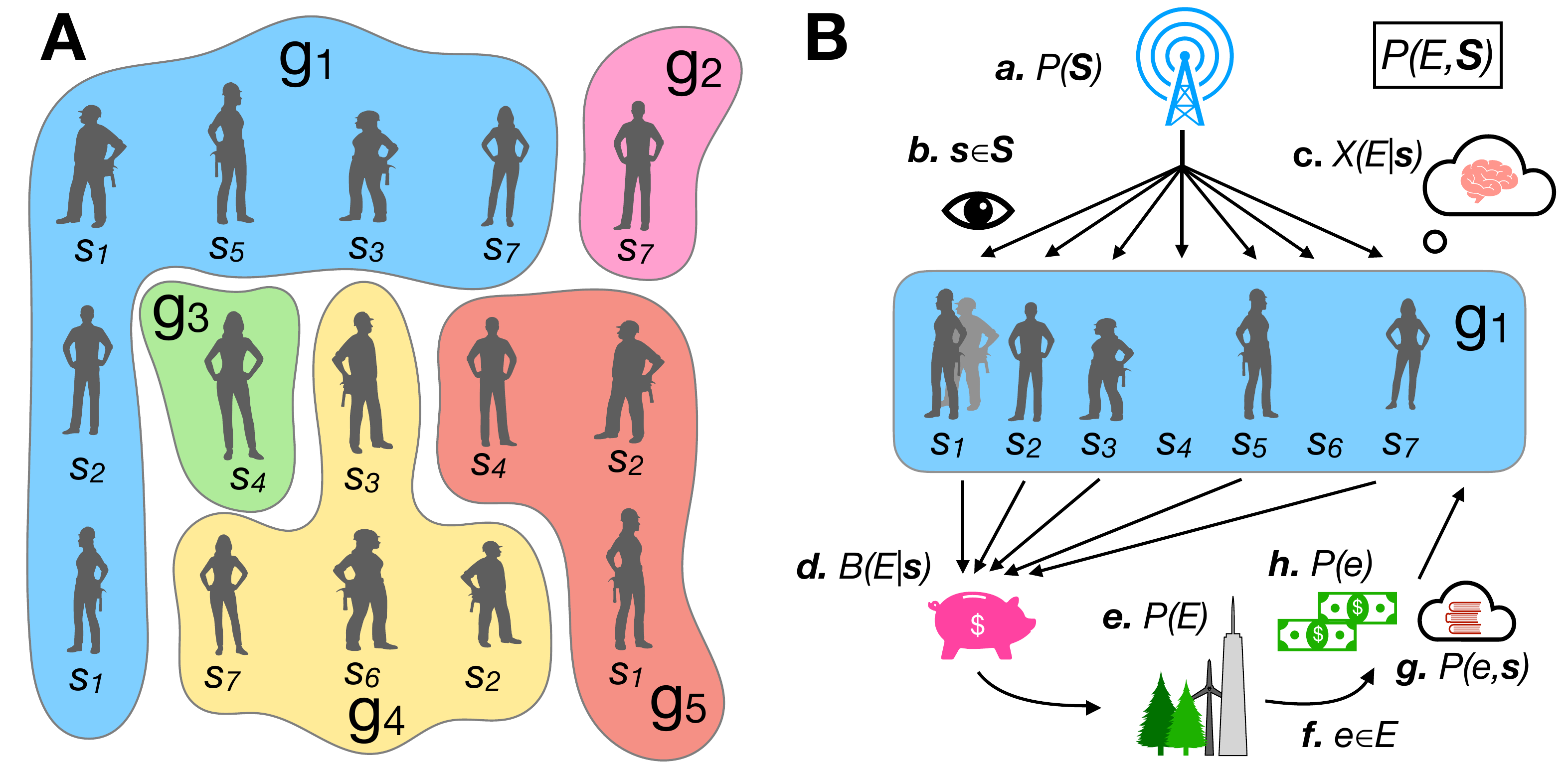}
    \caption{
    Groups of agents with different signals grow resources based on the information between their  signals and states of the environment.
    \textbf{A.} Groups, denoted $g$, are composed of an arbitrary number of agents.
    Each agent belongs to only one group and can observe  and contribute one signal to the group.
    A group contains $k_g$ unique signals. 
    \textbf{B.} At each time step, (a) the groups’s private channel outputs a signal $\boldsymbol{s}\in\boldsymbol{S}$ with probability $P(\boldsymbol{s})$. (b) Each member of the group observes their signal $s_j$ and (c) the group consults their collective belief for the conditional outcome probability of the environment, $X(E|\boldsymbol{s})$. (d) The agents make proportional resource allocations on all possible outcomes $B(E|\boldsymbol{s})$. (f and e). The true event $e\in E$ is observed in the environment with probability $P(e)$, and (g) the agents receive payouts proportional to the marginal probability of $e$. 
    }
    \label{fig1}
    \label{fig:my_label}
\end{figure*}

\noindent 
The coefficient of redundancy can have a positive or negative value, indicating different conditional relationships between the signals and environmental states.  When $R_P>0$, there is information across signals irrespective of environmental events. This means that signals are   $redundant$, and consequently there are diminished returns to pooling information as $I(E;\boldsymbol S)<\sum_jI(E;S_j)$.
Conversely, when $R(E;\boldsymbol S)=0$, the signals are statistically independent, and the benefits of pooling information increase linearly with the information of each signal on the environment but there is no synergy. Finally, when $R(E;\boldsymbol S)<0$, there is conditional dependence of the signals on the environment. This is called 
\textit{synergy} and corresponds to a superlinear benefit to pooling information in the number of agents, above and beyond the information contributed from each signal individually.

\subsubsection{Group formation and collective decision making}

We have now defined individual resource growth as a quantity of information and discussed how information can be aggregated across signals to express their synergy relative to states of the environment.
Now we can explore how agents with different signals can pool information together as coordinated groups, and access the synergy in their environment through collective decision-making. 

Consider the undirected hypergraph $H=(A,G)$ of vertices, $A$, and hyperedges $G$. 
We consider a discrete number of vertices, $A=\{a_1,a_2,\dots,a_N\}$, where $a_i$ identifies agent $i$. 
The set of hyperedges, $g\in G=\{1,2,\dots\}$, called groups, defines the number of cooperating collectives.
A hyperedge connects $1\leq N_g\leq N$ agents, and we assume that agents can only belong to a single group.
Therefore, by construction, $\sum_g N_g=N$ and the sum over all nodes of every hyperedge yields the number of agents in the population.
There exist two extremes of cooperation. 
First, when a single hyperedge spans every node, meaning all agents pool information in one  group.
In the limit of no cooperation, $N_g=1$ for all $g$, and no agents pool information.  In this case, the dynamics of the model are similar to previous work~\cite{kemp2023learning}.

Let $\boldsymbol S_g$ be the set of unique signals held by the agents of a group $g$ to be pooled, such that $\boldsymbol S_g\subseteq\boldsymbol S$.
The number of $cooperants$ is defined by the number of unique signals, $|\boldsymbol S_g|=k_g$, and is bounded by $1\leq k_g\leq l$.
When $k_g=l$ and the group has a complete signal, the collective can make maximally informed decisions.
Conversely, when $k_g<l$, the signal is considered $incomplete$, and the collective can only interpret and act on a subset of signals.
As we will see, the number of unique signals a collective can observe determines the amount of information they can access.

Now that we have defined how agents organize into groups of various sizes, we can discuss how agents pool their information to make collective decisions and grow their resources in dynamic environments. 
At every time step, a collective with access to all signal types observes a unique private signal $\boldsymbol s=\{s_1,\dots,s_l\}\in\boldsymbol S$.
Each agent then allocates its resources $r_i$ on events according to collective $g$'s allocation matrix $B(E|\boldsymbol s)$.
As the event $e$ is observed, the agent is rewarded with returns $w_e$ to the fraction of resources invested in $e$, $B(e|\boldsymbol s)$.
In the limit of many sequential investments $n$, the average growth rate of resources converges to

\begin{equation}
    \gamma=\frac{1}{n}\log\frac{r_n}{r_i}\approx\sum_{e,\boldsymbol s}P(e,\boldsymbol s)\log\big[B(e|\boldsymbol s)w_e\big],
\end{equation}

\noindent The optimal investment in the large $n$ limit is the conditional probability of the event given the signals, $B(e|\boldsymbol s)=P(e|\boldsymbol s)$.
When the rewards are ``fair", and $w_e=1/P(e)$, the optimal growth rate is given by the mutual information \cite{kelly1956new} defined in equation \ref{eqn1}, $\gamma=I(E;\boldsymbol S)$.

The typical collective may not have a complete signal, and instead may only observe and interpret a subset of all unique signals $\boldsymbol S_g$. 
Their optimal allocation, given by  $P(E|\boldsymbol S_g)$, then has mutual information 
$I(E;\boldsymbol S_g)\leq I(E;\boldsymbol S)$, with equality only if the omitted signals are completely redundant with present signals.
Unless there are redundant signals, an incomplete  group is guaranteed to have suboptimal information and growth rate.

Furthermore, agents may also not start out with perfect knowledge and must invest using their best estimate of the true environmental probability, $X(E|\boldsymbol S_g)\neq P(E|\boldsymbol S_g)$.
In this case, the collective's average growth will be submaximal by the number of signals and lack of information on signals, and is described by
\begin{equation}\label{kellyGroupGrowth}
    \gamma_{g}=I(E;\boldsymbol S_g)-\textrm E_{\boldsymbol s_g}\big(D_{KL}\big[P(E|\boldsymbol s_g)||X(E|\boldsymbol s_g)\big]\big),
\end{equation}

\noindent 
where $\textrm{E}_{\boldsymbol s_g}$ is the expectation value over the states of the group's signals, and $D_{KL}\big[P(E;\boldsymbol s_g)||X(E;\boldsymbol s_g)\big]=\sum_eP(e|\boldsymbol s_g)\log(P(e|\boldsymbol s_g)/X(e|\boldsymbol s_g))\geq0$ is the Kullback-Leibler divergence, an information measure expressing how similar the distributions are in their inputs.
This result shows that collectives with both a better model as reflected by the first term, a better characterization of the model and its various synergies by the second, and a more complete signal will experience higher growth rates. Furthermore, $\gamma_g<\gamma$ unless $g$ is the full set of signals, so it is typically valuable to add more signals to the group. This setup is illustrated in Figure \ref{fig1}.

\subsection{Maximum synergy principle and optimal growth}

These results introduce important considerations for how  collective innovation and growth determine strategies for group formation.
In theories of cooperation such as kin selection \cite{eberhard1975evolution} and scalar stress \cite{johnson1982organizational}, group formation is advantaged by member relatedness and disadvantaged by unfamiliarity. 
This is intuitive in many situations, as agents are more likely to cooperate when they are more certain others will reciprocate \cite{nowak2006five}, and cooperating with similar agents minimizes this uncertainty. Equation \ref{kellyGroupGrowth}  counters this intuition by defining an explicit benefit to cooperating with dissimilar agents across heterogeneous, complementary skills and information.
Specifically, a group with more synergistic signals, as defined through the conditional dependence of their decisions on states of the environment, will experience higher growth. So, even if there are additional coordination costs for more heterogeneous agents, there is now a possiblity that cooperation will emerge as there are also greater informational benefits, formalizing intuitive ideas about the value of diversity~\cite{page_difference_2007}.

The beneficial contribution of synergy to the growth rate of resources provides an important input to models of random multiplicative growth, such as those commonly used to study wealth dynamics and mathematical finance. 
In its simplest form, the stochastic growth rate in such models is characterized by its first two temporal moments. The average over time, $\eta$, and the resource temporal standard deviation (volatility), $\sigma$, combine under It\^{o} integration to give actual growth rate $\gamma=\eta-\sigma^2/2$. Maximizing this growth rate (as a positive quantity) entails maximizing $\eta$ and minimizing $\sigma$, which at the individual agent level can be achieved by (Bayesian) learning over time~\cite{kemp2023learning}.

At the population level, it has been proposed that pooling resources in groups would naturally emerge as a means to reduce $\sigma$, when growth rate fluctuations are independent across agents, and thus maximize $\gamma$~\cite{fant2022stable, peters2022ergodicity}.

Our results introduce a different possibility of cooperation, through pooling information in structured groups, that maximizes $\eta$ (and $\gamma$) through synergy effects. Thus, to maximize $\gamma$, agents should pool information with a most diverse set of collaborators possible to access the most mutual synergy viz. the environment.
This \textit{maximum synergy principle} defines the benefit of intelligent collective behavior in complex environments where there are agent level limitations to knowing the environment fully and where mechanisms of the division of labor and knowledge are favored. This principle is general and applies across levels of cooperation, whether it be individuals matching skills to form groups, or specialized groups organizing into more complex collectives \cite{bettencourt2009rules}, all the way to large scale societies.

\begin{figure}
    \centering
    \includegraphics[width=.45\textwidth]{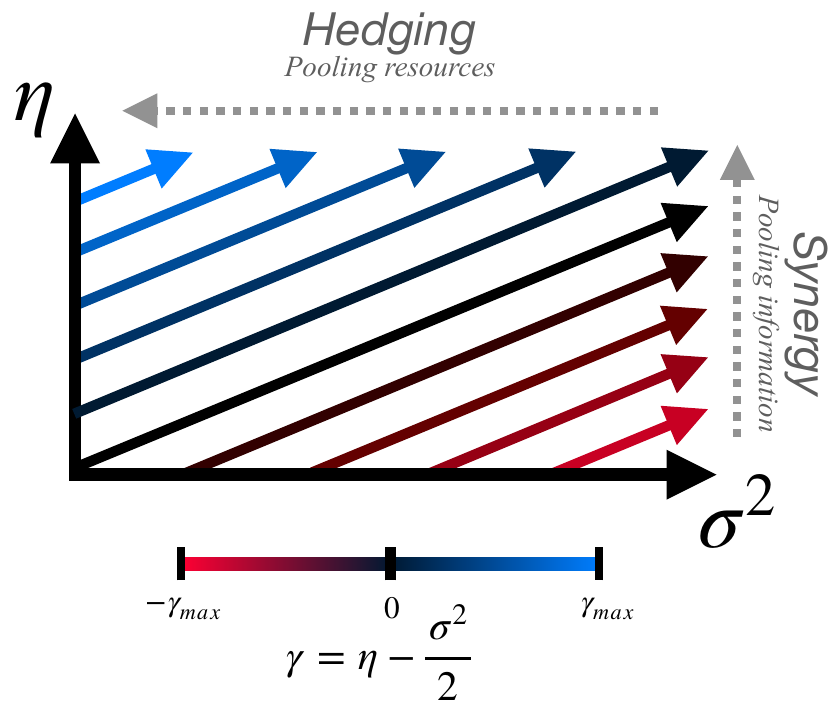}
    \caption{
    Complementary strategies for increasing the long-term growth rate of resources from the environment in stochastic growth models. 
    Pooling resources can reduce the volatility through a hedging strategy while pooling information creates synergy to increase average growth rates.
    The lines represent contours of constant average growth rates $\gamma$. 
    }
    \label{fig2}
    \label{fig:my_label}
\end{figure}

Generally, these two strategies-- information synergy versus resource pooling under independence-- are distinct modes of cooperation over which groups can maximize $\gamma$, as demonstrated in Figure \ref{fig2}.

As we will see later, the decision of who to cooperate with is not trivial, as different combinations of signals may yield varying synergies.
This means that under constraints of group size, such as from cooperation costs per connection, groups satisfying the maximum synergy principle must intelligently select which signals and agents to integrate, and which to exclude as redundant. 

Furthermore, collectives may not \textit{a priori} know the optimal allocation strategy that leverages the synergy available to their signals, meaning that intelligent collective behavior must itself be learned over time and by exploring the best possible matchings. We will now develop the dynamics of how a group can organize itself optimally so as to maximize its synergy.

\subsubsection{Synergy maximization through Bayesian inference}

Bayesian learning is the optimal strategy to incorporate new information from observed events into the estimate of conditional probabilities, such as those of environmental states given agents' signals \cite{kemp2023learning}. Agents can also learn the synergy embedded in their environment in groups by collectively weighing their conditional observations across their individual signals.  A group wanting to maximize their synergy must then update their conditional relationship through a Bayesian inference process
\begin{equation}
X_n(e|\boldsymbol s)=AP(\boldsymbol s_n|e_n)X(e_n)=\bigg[\prod_{i=1}^n\frac{P(\boldsymbol s_i|e_i)}{P(\boldsymbol s_i)}\bigg]X(e),
\end{equation}
\noindent where the normalization $A=(\int de_nP(\boldsymbol s_n|e_n)X(e_n))^{-1}$.
We take the prior probability, $X(e_1) = X(e)$, because we are assuming that the environment is stationary or at least slowly changing relative to groups’ learning rates.

Bayesian inference converges $X(E|\boldsymbol S)\rightarrow P(E|\boldsymbol S)$ over time, decreasing the information divergence, and maximizing synergy and average growth.  For groups with incomplete signals, the information acquired through learning is still bounded by what is available in the incomplete signal space.

We have thus far defined  collective growth in terms of information synergy, and shown how agents can learn as a collective to increase their growth rate over time.
We will now illustrate these general results using a model based on logic circuits.

\section{Modeling Synergy with Logic Circuits}

Logic circuits have been used extensively as models for synergistic interactions \cite{bettencourt2009rules,schneidman2003synergy,bettencourt2007functional}. This  is because their outputs are predicted by combinations of inputs, much like events are predicted by combinations of signals. Among other logic circuits (like AND or OR), the XOR gate is unique in that information between inputs and outputs only exists as synergy across all inputs \cite{jansma2022higher}; no individual input has mutual information with the output. 

In the following, we will show how modifying the XOR gate relaxes this condition, such that information exists for any input and scales on average with the number of cooperating signals.  Similar to \cite{kemp2023learning}, while this model will be used to study synergy in a  simplified setting, the theory is defined for general dynamical environments. 

\subsection{The Uniform XOR Gate}

Consider the space of statistically independent binary signals $s_j\in 0,1$, such that a sample set  $\boldsymbol s$ has uniform probability $P(\boldsymbol s)=2^{-l}$.
We assign each input $\boldsymbol s$ a binary event, $e\in 0,1$, using the generalized XOR rule, $e=M_2(\boldsymbol s)\equiv\big[\sum_{j=1}^ls_j \big](\textrm{mod 2})$ with binomial probability $p_{\boldsymbol{s}}$. 
From the sets of sampled signals, $\boldsymbol s$, and  binomial coefficients $\boldsymbol p=\{p_{\boldsymbol{s}}\}$, we can define this generalized  XOR circuit as a joint distribution on signals and events as

\begin{equation}\label{pxorInfo}
\begin{split}
    P(E,\boldsymbol S|\boldsymbol{p})\equiv&f(\boldsymbol p,l)=\frac{1}{2^l}\prod_{\boldsymbol s}(p_{\boldsymbol s})^{M_2(\boldsymbol s)}(1-p_{\boldsymbol s})^{1-M_2(\boldsymbol s)}.
\end{split}
\end{equation}

\noindent This distribution is called the uniform XOR (UXOR). It performs a unique, $l$ dimensional XOR gate on each input $\boldsymbol s$ with probability $p_{\boldsymbol s}$.  In the case  where $p_{\boldsymbol s}=1$ for all input permutations, this circuit behaves deterministically like an XOR gate, and the complete group has 1 bit of information.  In the limit of $p_{\boldsymbol s}=.5$, this no longer models a logic gate as the output is uncorrelated to the inputs. 
The truth table of this circuit is shown in Figure \ref{fig:fig2}A for an environment with two signals. 


\subsubsection{Information scaling in the UXOR environment}

With this explicit choice of distribution, we can explore quantities of information that will define a group's growth process.
For simplicity, we choose a uniform prior for the distribution of $\boldsymbol p$, but in principle any prior distribution is admissible.
The information available in the environment measures the maximum average growth rate a group with a complete signal can experience.
When averaged over all configurations of $\boldsymbol p$, the information is given by $I(E;\boldsymbol{S})=\log2-1/2\approx .28$ bits (App. \ref{sec:mi_under_complete_cooperation}).

For groups with incomplete signals (when  $k_g<l$), we compute the information by marginalizing equation \ref{pxorInfo} over the $\lambda_g=l-k_g$ signals unavailable to the group.  The procedure for marginalization is defined in App. \ref{sec:mi_under_complete_cooperation}, but in general, 
marginalization of one signal halves the size of the parameter space $\boldsymbol p$ that describes the distribution.
The average information for an incomplete signal is approximately (App. \ref{sec:mi_under_incomplete_cooperation})

\begin{equation}\label{eq:mi_scaling_estimate_lln}
    I(E;\boldsymbol S_{g}|\boldsymbol p)\approx 2^{-\lambda}\left(\log2 - \frac{1}{2}\right).
\end{equation}

\noindent Average information increases exponentially, $\sim 2^k$, as more signals are included. The mutual information of the complete signal is independent of the number of signals, so the information of a single signal must converge to zero in the limit of large $l$.

\begin{figure}[h]
    \includegraphics[width=.45\textwidth]{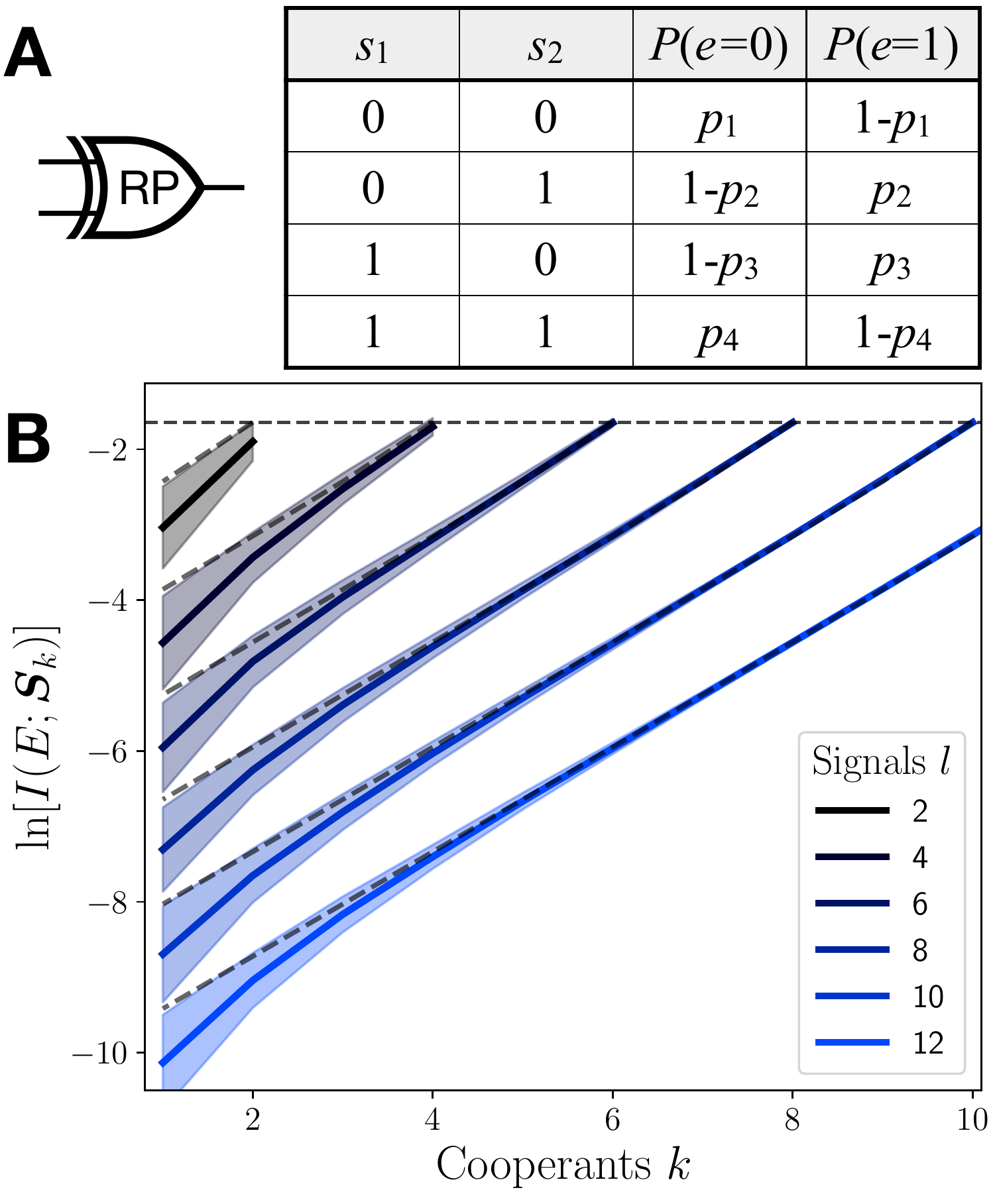}
    \caption{The UXOR model provides an environment for exploring synergy across groups of arbitrary size.
    \textbf{A.} The UXOR circuit, demonstrated by the modified XOR symbol, and its truth table for $l=2$.
    \textbf{B.} The information of a circuit of size $l$ scales exponentially in cooperating signals, $k$. 
    }
    \label{fig:fig2}
\end{figure}

The exponential scaling of the information with the number of cooperants is demonstrated in Figure \ref{fig:fig2}B, as lines on a logarithmic scale for environments of increasing $l$.  The curves are computed by Monte Carlo sampling circuits for $l$ signals by measuring the information after $\lambda=l-k$ marginalizations. 

\subsubsection{Growth and group learning}

Until now we have explored the mean behavior of this environment subject to a uniform prior. 
In general, collectives do not have perfect information on a single prior.
In this case, their inaccurate guess for the set of binomial coefficients is parameterized by $\boldsymbol x_g\equiv \{x_{\boldsymbol{s}_g}\}$, indexed by the signals available to the group $
\boldsymbol s_g\in \boldsymbol S_g$, and the collective's likelihood model becomes $X(e|\boldsymbol s_g)=f(\boldsymbol x_g,k_g)$.
The information divergence term of equation  \ref{kellyGroupGrowth} becomes the divergence between $f(\boldsymbol x_g,k_g)$ and $f(\boldsymbol p_g,k_g)$, where $\boldsymbol{p}$ has been projected into the subspace spanned by $\boldsymbol S_g$, averaged over all signals $\textrm E_{\boldsymbol s_g}[D_{KL}]=\langle p_{\boldsymbol s_g}\log(p_{\boldsymbol s_g}/x_{\boldsymbol s_g})+(1-p_{\boldsymbol s_g})\log[(1-p_{\boldsymbol s_g}/(1-x_{\boldsymbol s_g})]\rangle$ here angle brackets denote sample averages over the binomial values.
Subtracting the mutual information by this term yields the growth rate under imperfect, incomplete group information.

\begin{equation}
\gamma_g=\big\langle p_{\boldsymbol s_g}\log x_{\boldsymbol s_g}+(1-p_{\boldsymbol s_g})\log\big(1-x_{\boldsymbol s_g}\big)\big\rangle+\log2,
\end{equation}

We have so far described growth rate dynamics under a stationary $\boldsymbol x_g$.
To illustrate growth dynamics under group learning, we turn to the Latent Dirichlet Allocation (LDA) model. Through a categorical description of pairs of events and signals, agents experience average dynamics to $\boldsymbol x_g$ in the limit of high sampling rate $\omega=n/t\gg1$

\begin{equation}\label{lda}
\boldsymbol x_{g}(t)=\frac{\boldsymbol p_{ g}t/2\kappa+\boldsymbol x_{g}}{1+t/2\kappa},
\end{equation}

\noindent where $\kappa$ defines the Bayesian update time.
The details of LDA are given in Ref.~\cite{kemp2023learning} and provide parametric 

\noindent dynamics that converge to full information as a power law in time, in stationary environments.

\begin{figure*}
    \includegraphics[width=.95\textwidth]    {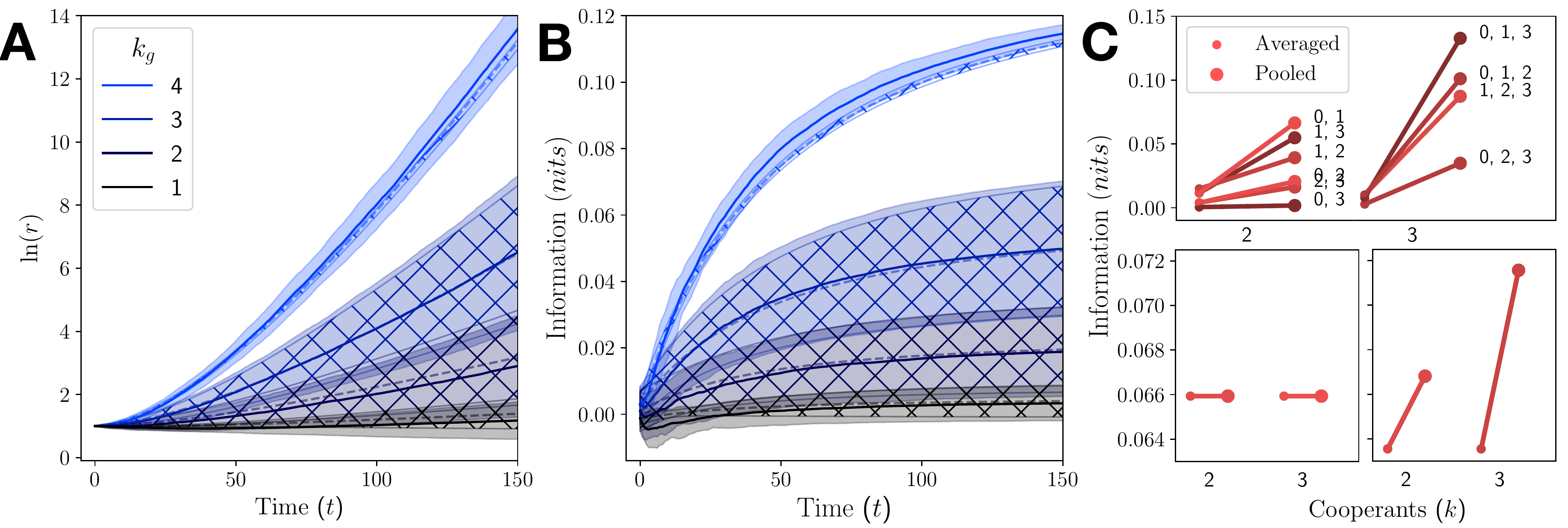}
    \centering
    \caption{Groups learning an $l=4$ environment using  more unique signals acquire more resources and information, but combinations of signals have unique amounts of information.
    \textbf{A}. 
    Temporal resource trajectories, grouped by number of unique signals in the corresponding group show that  growth increases with the number of signals.
    \textbf{B}. Groups with more signals can gather more information from the environment.
    There is high variability when $k_g<l$, as different combinations of signals access different amounts of information.
    \textbf{C}. \textit{Top} For $k_g=2,3$, the synergy benefits of a parameter configuration are given by the difference between the information when averaged (small dot) and pooled (large dot).
    \textit{Bottom} Parameter values exist where no signal combinations hold synergy (left) and synergy is equivalent across signal combinations (right). 
    }
    \label{fig:fig4}
\end{figure*}

To study the dynamics of resources in the UXOR environment, we simulated agent investments in a Monte Carlo sampled environment.
We randomly assigned $N=2000$ agents signals in an $l=4$ environment, then randomly assigned them to groups sized $l\leq N_g\leq 11$. 
This results in an ensemble of groups with cooperants $1\leq k_g\leq 4$. 
We reveal Bernoulli-sampled signals to the groups, whose agents make collective decisions on which events to allocate resources.  
For each group, we track the resources of a representative agent, informed by the group, investing their individual resources through time.

Figure \ref{fig:fig4} illustrates the results of this simulation.
In subfigures A and B, the Monte Carlo simulated means are shown as solid lines, with 95\% Confidence Interval (CI) shaded regions. 
Theoretical means are computed from the initial population configuration using equation \ref{lda}, plotted as dashed lines, with hash-filled uncertainty regions.
The more unique signals a group can access increases, the more they can learn, and the more resources they acquire over time.
A high signal-to-noise ratio when $k_g=1,2$ causes growth rates lower than the theoretical mean, and cumulatively fewer resources over time.  

\subsubsection{Constrained intelligent group formation}

For the groups with $k_g<4$ (incomplete signals) there is significantly higher variance in both information and resources compared to $k_g=4$.
This is attributed to differences in synergy between groups with different combinations of signals of order $k$. 
This illustrates a general feature of the maximal synergy principal; that signal combinations with higher conditional dependence on the environment will have higher synergy and experience higher growth rates than other combinations. 
Figure \ref{fig:fig4} demonstrates the synergy effects across different combinations of signals. 
For each group of size $k$, the left, smaller dot indicates the amount of information each signal has averaged over the signals present. 
The right, larger dot indicates the total information the combination of signals has when pooled.
The difference between the two dots gives the amount of synergy.
We see, for example, that even though signals 0 and 3 have less information than signal 2, both signals have higher synergy effects when pooled with 1 individually, as indicated by their crossover with the 1, 2 line. 
For a group aggregator, not only does this mean that signal choice is nontrivial, but also that individual information is not generally a good indicator of synergy benefits that can be realized when pooled.  
This result reinforces the complexity that fulfilling the maximum synergy principle entails, as one must understand signal complementarities for a given model of the environment to all orders, a likely costly process.

As demonstrated by the bottom plots in Figure 4C, through a smart selection of $\boldsymbol p$, we can also design special environments such as where either no synergy is present, or where there are uniform benefits of synergy across combinations of signals. 
The procedure for constructing environments with specific synergy profiles will developed in future work.


\section{Discussion}

In this paper, we developed a novel mechanism of cooperation among heterogeneous agents that use shared information to grow resources in noisy environments. 
We derived the benefits of cooperation in terms of synergy gained by pooling information across agents' unique signals.
This motivates the principle of maximum synergy, whereby a group's aggregate growth is optimized when that group maximizes the synergy of its members relative to a statistical environment. We proposed this principle as a complementary avenue to cooperation compared to the reduction of  volatility through resource pooling in multiplicative growth models.
We then showed that a group with no \textit{a priori} knowledge of its potential synergy can learn it through Bayesian inference.
We illustrated these principles using a model of a high-dimensional probabilistic logic gate and showed that, on average, group synergy scales superlinearly (exponentially) with the number of unique signals in the group.
We also illustrated the challenge faced by groups under constraints to size to pick not just unique signals but also admit new group members as additional signals that maximize their potential collective synergy.

These results formalize  several insights into the causes and benefits of cooperation. First, the properties of information allow us to consider how the limits to human effort and ability motivate group formation. Specialization through learning or adaptation is costly in terms of time and resources, motivating a division of labor to fully learn and maximize productivity across disparate but synergistic agents  \cite{hume2003treatise}.


Second, these results motivate analyses of how information and resource pooling strategies affect different levels of selection within an organizational hierarchy. 
Effective resource pooling relies on uncorrelated fluctuations across participants, which is not possible when agents are making coordinated decisions across correlated signals.
We therefore expect information and resource pooling strategies to create tradeoffs in group formation, and apply to different environmental features and levels of selection.

Groups lacking informational complementarities (because they are homogeneous) operating in variable environments should pool resources. This may apply to people in insurance pools, or independent economic sectors within a common population, such as a city or nation. Conversely, groups in complex environments made up of agents with complementary knowledge, such as within a firm or innovation ecology, should engage in information pooling and skills specialization to maximize their production whenever the variability of the environment and costs of cooperation are sufficiently low.  

Parsing out these modes of cooperation becomes more important when considering how groups respond to changing environmental or social conditions.
As new environmental conditions emerge, such as new industries, the distribution of synergy across different group configurations will also change, selecting for different group compositions and skills combinations. This has the interesting implication that new knowledge (science, technology, institutional change) should be disruptive of established social and economic structures because it enables new synergies. 
This also has implications for natural ecosystems~\cite{walther2010community} where changing environmental conditions and variability, such as via climate change, may alters their structures.

Third, the framework developed here describes a general approach to describing interaction dynamics in many fields. 
The conditional probabilities $P(e|\boldsymbol{s})$ capture the general structure of information between populations and their environment. Through synergy, that information becomes encoded in how groups form and are structured, and which sets of coordinated behaviors produce beneficial or detrimental behaviors across agents.
By tracing over states of $E$, averaging over (stationary) environments, we can produce a set of rules for (average) rewards associated with agents' perceptions and actions.  This shows how general conditional probabilities of choices and behaviors in given environments may underlie particular games and other phenomenological agent interaction rules~\cite{autor2014skills}. Furthermore, because conditional distributions are general and multi-dimensional they also provide natural models of higher order interactions expressing large groups' synergy, such as reciprocal cooperation and the emergence of culture as shared knowledge and behavior~\cite{queller1985kinship}. In summary, the formal properties of information, made explicit over group structures and time, provide the theoretical basis for a broad class of agent interaction models found throughout the social and ecological sciences. This includes the formation of complex societies made up of diverse cooperating agents in situations where large scale synergy becomes possible.

This work is supported by the Mansueto Institute for Urban Innovation and the Department of Physics at the University of Chicago and by a Na- tional Science Foundation Graduate Research Fellowship (Grant No. DGE 1746045 to JTK), and by the National Science Foundation, through the Center for the Physics of Biological Function (PHY-1734030), as well as the National Institutes of Health BRAIN initiaitive (R01EB026943) to AGK.

\begin{widetext}
\section{Appendix}
\subsection{Information Aggregation}\label{infoagg}

Consider a target statistical variable $E$ (environment), that we wish to predict using $l$ other variables (signals) $\boldsymbol S=\{S_1,\dots,S_l\}$.
The mutual information between each signal $S_i$ separately and $E$ is given by
\cite{cover1991information}

\begin{equation}
    I(E;S_i)=H(E)-H(E|S_i)=-\frac{\Delta H(E)}{\Delta S_i}.
\end{equation}
\noindent where $H(E)$ is the Shannon entropy of $E$, and the variation measures the difference in entropy of the event when conditioned on the signal. From the rules of information aggregation, this expression generalizes to information across every added signal \cite{bettencourt2008identification}.
The mutual information between the event and the set of several signals is given by
\begin{equation}
\begin{split}\label{entropyseries}
    I(E;\boldsymbol S)=&-\sum_{i=1}^l\frac{\Delta H(E)}{\Delta S_i}
    -\sum_{i>j=1}^l\frac{\Delta^2H(E)}{\Delta S_i\Delta S_j}
    -\dots -\frac{\Delta^lH(E)}{\Delta S_1\dots\Delta S_l}. 
\end{split}
\end{equation}

The first term of this expansion is just a sum over the mutual information of each individual signal and the environment. The goal of this section is to show that the inclusion of each new signal introduces a coefficient of redundancy of progressively higher order. The first term is 

\begin{equation}
    \begin{split}
        \frac{\Delta ^2 H(E)}{\Delta S_i\Delta S_j}=&H(S_i)-H(S_i|E)+H(S_j)-H(S_j|E)
        -H(S_i,S_j)+H(S_i,S_j|E)\\
        =&H(S_i)-H(S_i|E)-H(S_i|S_2)+H(S_i|S_j,E)\\
        =&I(S_i;S_j)-I(S_i;S_j|E)\equiv R(E;S_i;S_j),
    \end{split}
\end{equation}

\noindent where we used the identity $H(A,B)=H(A|B)+H(B)$.
We denote $R$ as the coefficient of redundancy, which measures the difference in mutual information between the variables, $I(\boldsymbol{S})\equiv I(S_1;\dots,S_k)$, and the mutual information of the variables conditioned on $E$, $I(\boldsymbol{S}|E)$.
When $I(S_i;S_j)<I(S_i;S_j|R)$, the signals contain less mutual information in the absence of the event (we gain information by considering the event), and  $R(S_i;S_j;X)<0$. In this case agents experience a positive benefit from pooling information, which we call synergy.

To demonstrate this effect to higher orders in goups of signals, we perform a similar calculation for a three-signal interaction. 
\begin{equation}
    \begin{split}
        \frac{\Delta ^3 H(E)}{\Delta S_i\Delta S_j\Delta S_k}=&H(E)-H(E|\{S_i,S_j,S_k\})\\
        =&H(S_i)+H(S_j)+H(S_k)-H(S_i|E)
        -H(S_j|E)-H(S_k|E)-H(S_i,S_j)-H(S_j,S_k)\\
        &-H(S_k,S_i)+H(S_i,S_j|E)
        +H(S_j,S_k|E)+H(S_k,S_i|E)
        +H(S_i,S_j,S_k)-H(S_i,S_j,S_k|E)\\
        =&H(S_i,S_j,S_k)-H(S_i|S_j)-H(S_j|S_k)
        -H(S_k|S_i)+H(S_i|S_j,E)+H(S_j|S_k,E)\\
        &+H(S_k|S_i,E)-H(S_i,S_j,S_k|E)\\
        =&I(S_i;S_j;S_k)-I(S_i;S_j;S_k|E)\equiv R(E;S_i,S_j,S_k).
    \end{split}
\end{equation}

\noindent We see that an analogous redundancy coefficient arises in three dimensions.
This can generally be retrieved for arbitrary number of dimensions through a similar iterative procedure.  We refer to the sum of these moments collectively as the redundancy of the joint distribution, denoted $R_P$ \cite{bettencourt2008identification},

\begin{equation}\label{completesynergy}
\begin{split}
    R_P\equiv&
    -\sum_{i>j=1}^l\frac{\Delta^2H(E)}{\Delta S_i\Delta S_j}-\dots -\frac{\Delta^lH(E)}{\Delta S_1\dots\Delta S_l} \\
    =&\sum_{i>j=0}^l\big[I(S_i;S_j)-I(S_i;S_j|E)\big]+\dots
    +I(S_1;\dots;S_l)-I(S_1,\dots,S_l|E)
    \end{split}.
\end{equation}

\noindent Note that redundancies of lower order than cardinality of the signal space must be computed over every combination of signals. For example, when $l=3$, there are three second order redundancy terms.

This expansion generally defines the benefits to cooperation over increasingly higher orders of cooperation (number of signals). This expression can be used to compute the relative strengths of the various orders of interaction for any set of signals and environmental variables, given their conditional distributions. 

\subsection{Kelly Growth rate}

Consider an environment with events conditionally dependent on signals characterized by a joint distribution $P(E,\boldsymbol{S})$ for event $E$ and $l$ signals $S$.
Consider a cooperative Kelly investment scheme whereby each participant, agent $i$, witnesses signal $s_i\in S_i$, and informs the collective how to invest their shared resources $r$.
The mechanics of pooling resources and collectively investing will be discussed below.
Kelly's formalism can be adapted by expanding the environmental probability to contain $l$ signals, $P(E,S)\rightarrow P(E,\boldsymbol{S})$, as can the betting matrix $X(E|S)\rightarrow X(E|\boldsymbol{S})$, where $\boldsymbol{S}=\{S_1,\dots,S_l\}$.
When odds are fair, the Kelly growth rate is given by the returns to each investment, averaged over the probability of that signal, event pair 

\begin{equation}
    G=\sum_{e,\boldsymbol{s}}^{E,\boldsymbol{S}}p(e,\boldsymbol{s})\log\frac{x(e|\boldsymbol{s})}{p(e)}.
\end{equation}

We expand this equation by inserting $p(e,\boldsymbol{s})$ into the numerator and denominator of the log

\begin{equation}
    G=\sum_{e,\boldsymbol{s}}^{E,\boldsymbol{s}}p(e,\boldsymbol{s}) \bigg[ \log \frac{p(e,\boldsymbol{s})}{p(\boldsymbol{s})p(e)} -\log  \frac{p(e|\boldsymbol{s})}{x(e|\boldsymbol{s})} \bigg].
\end{equation}

\noindent These two terms can be simply expressed as $G=I(E;\boldsymbol{S})-\textrm{E}_{\boldsymbol{s}}\big[D_{KL}\big(P(E|\boldsymbol s)||X(E|\boldsymbol s)\big)\big]$, similar to previous work, but we can decomposes this equation in terms of redundant information across the signals using equations \ref{entropyseries} and \ref{completesynergy}. 

\subsection{Information for UXOR circuits}\label{sec:mi_under_complete_cooperation}

Here we compute  $I(E;\boldsymbol{S})$ for  the UXOR logic circuit. This represents the information that a group of agents with $l$ distinct signals have about the output $E$ of the probabilistic gate, averaged over all configurations of the gate for a uniformly distributed prior. Because the signals $s_i$ are independent Bernoulli trials with probability $1/2$,  
\begin{equation*}
    I(E;\boldsymbol S) = \sum_{e,\boldsymbol s} P(e,\boldsymbol s) \log \frac{P(e,\boldsymbol s)}{P(e)P(\boldsymbol s)} = \frac{1}{2^l} \sum_{e, \boldsymbol s} P(e|\boldsymbol s) \log \frac{P(e|\boldsymbol s)}{P(e)}\,.
\end{equation*}
Then, using the fact that the output $E$ is also a Bernoulli variable, $P(e=1|\boldsymbol s) = 1-P(e=0|\boldsymbol s)$, and
\begin{equation}\label{eq:complete_cooperation_MI}
    I_l = I(E;\boldsymbol S) = \frac{1}{2^l} \sum_{\boldsymbol s} g(P(e=0|\boldsymbol s)) - g(P(e=0))\,,
\end{equation}
where $g$ is a function representing application of the  UXOR gate and is defined over binomial parameters $x$ as
\begin{equation*}
    g(x) = x \log x - (1-x) \log (1-x)\,.
\end{equation*}
The number of terms in \eqref{eq:complete_cooperation_MI} grows exponentially with $l$ and quickly becomes large. When it is sufficiently large, the sum can be approximated by an average. In particular, for uniformly distributed $P(e=0|\boldsymbol{s})$,
\begin{equation*}
    \frac{1}{2^l}\sum_{\boldsymbol s}g(P(e=0|\boldsymbol s)) \approx \langle g(x) \rangle_{x\sim U(0,1)}.
\end{equation*}

\noindent This expectation can be analytically evaluated. With this, \eqref{eq:complete_cooperation_MI} gives
\begin{equation*}
I_l = \langle g(x)\rangle_{x\sim U(0,1)} - g(1/2) = \log 2 - \frac{1}{2}\,.
\end{equation*}

\subsection{Mutual Information for incomplete signal sets }\label{sec:mi_under_incomplete_cooperation}



Here, we demonstrate that the information the collective has about the gate output scales exponentially with respect to the number of cooperants, $k$, as is depicted in Figure \ref{fig:fig2}. Following the previous section, the introduction of the function $g(x) = x\log x + (1-x)\log(1-x)$ simplifies the expression for mutual information
\begin{equation}\label{eq:incomplete_cooperation_MI}
    I_k = I(E;\boldsymbol S) = \frac{1}{2^k} \sum_{\boldsymbol s} g(P(e=0|\boldsymbol s)) - g(P(e=0))\,.
\end{equation}
As before, this sum can be interpreted as an average over the uniform distribution when the number of terms is large. Here, the removal of parts of the signal changes the distribution of parameters, so the measure that approximates this sum also changes. 
We call this new measure $P_k$. Furthermore, whereas in the main text, the subscript of $\boldsymbol S_g$ denoted the signals of group $g$, here the subscript of $\boldsymbol S_k$ will denote the signal set of cardinality $k$ to be marginalized. With this new notation, the first term in \eqref{eq:incomplete_cooperation_MI} may be written approximately as:
\begin{equation*}
    \frac{1}{2^k}\sum_{\boldsymbol s_k}g(P(e=0|\boldsymbol s_k)) \approx \langle g(x) \rangle_{x\sim P_k}.
\end{equation*}
To compute $P_k$ for $k<l$, we need to calculate how the probability of $E$ conditional on the remaining signals $\boldsymbol s_k$ changes under the removal of the $k^{\text{th}}$ signal. 
For this model, 
\begin{equation}\label{marginalsk}
    P(e|\boldsymbol s_{k-1}) = \frac{1}{2}\left(P(e|\boldsymbol s_{k-1},s_k=0) + P(e|\boldsymbol s_{k-1},s_k=1)\right).
\end{equation}
By iterating this sum, we reduce the number of parameters required to describe $P_k(k)$, which in the main text are given by the set of binomial parameters $\boldsymbol p$.
Additionally, the distribution $P_k(x)$ becomes increasingly narrow, centered around 1/2, which is the mean of all probabilities $P(e|\boldsymbol s)$. Parameterizing $P_k(x)$ by its moments allows us to directly compute the mutual information. The moment expansion of the distribution is given by 
\begin{equation*}
    I_k \approx \langle g(x) \rangle_{P_k} - f\left(\frac{1}{2}\right) = \sum_{a=0}^\infty\frac{1}{a!}f^{(a)}(x_0)\langle (x-x_0)^a \rangle_{P_k} - f\left(\frac{1}{2}\right).
\end{equation*}

\noindent Using standard arguments, which we provide in the following section, these moments approximately scale like
\begin{equation} \label{eq:moment_scaling_approximation}
    m_{k-1}^{(a)} = \left\langle\left(x - \frac{1}{2}\right)^a\right\rangle_{P_{k-1}} \approx \frac{m_{k}^{(a)}}{2^{a/2}}\,,\qquad m_l^{(a)} = \frac{1}{(a+1) 2^a}\,,
\end{equation}
\noindent which is related to the onset of central limit theorem behavior. 
This provides us with a heuristic explanation for why $I_k$ scales as $1/2^{\lambda}$. 
After only a few cooperants are removed, higher order terms in the expansion (with order denoted by $a$) quickly die away, leaving only the second-order term
\begin{equation*}
    I_k \approx  \sum_{n=0}^\infty\frac{f^{(a)}(1/2)}{a!} m_k^{(a)} \to 2 m_k^{(2)} + O((m_k^{(2)})^2)\,.
\end{equation*}
Once this occurs, we can see plainly that between each marginalization the mutual information reduces by half,
\begin{equation*}
    \frac{I_{k}}{I_{k+1}} \to \frac{1}{2}.
\end{equation*}

While this explanation gives approximately the correct scaling behavior, it does not admit a good estimate of $I_k$ near full cooperation, since there the higher-order terms in the expansion are not small. To explicitly include these terms, we need all derivatives of $f$, evaluated at $x=1/2$
\begin{equation*}
    f^{(a)}(1/2) = \frac{(-1)^a2^a a!}{a(a-1)}.
\end{equation*}

\noindent Inserting these derivatives and the approximation \eqref{eq:moment_scaling_approximation} for moments of $P_k$ gives an approximation of $I_k$ as a series. Then, evaluating this series analytically yields a closed form expression.
\begin{align*}
    I_k &= \sum_{a=0}\frac{f^{(a)}(1/2)}{a!} m_k^{(a)} - g(1/2) \approx \sum_{a=1}^\infty \frac{(2a + 1)!}{(2a-1)!} \frac{1}{2^{\lambda a}} \\
        &= \frac{1}{2}\left[(2^{-\lambda/2} + 2^{\lambda/2})\, \text{arctanh}(2^{-\lambda/2}) + \log(1-2^{-\lambda}) - 1\right]\,, \qquad \lambda > 0
\end{align*}
This expression gives a good approximation for $I_k$ in the small $\lambda$ regime and also captures the scaling in the intermediate regime. For $\lambda\to\infty$, an estimate of the asymptotic behavior is given by setting $z = 2^{-\lambda/2}$ and Taylor expanding around $0$.
\begin{align*}
    \frac{1}{2}\left[(z + z^{-1})\, \text{arctanh}(z) + \log(1-z^2) - 1\right] &\to \frac{1}{2} \left[(z + z^{-1})(z + \frac{1}{3}z^3 + O(z^4)) - z^2 + O(z^4) - 1\right] \\ 
    &=\frac{1}{6} z^{2} + O(z^4)
\end{align*}

Because $z=2^{-\lambda/2}$, the quadratic leading term agrees with the observed exponential scaling $I_k \propto 2^{-\lambda}$. Although the scaling prediction is correct, there is not a regime where this high-$\lambda$ expression consistently estimates $I_k$ in its actual value. The essential reason is that as $k$ decreases, the number of terms in the sum over remaining signal states decreases, and the approximation of that sum as an average over $P_k(x)$ begins to break down. This manifests as an error in the overall scale of the estimate, but not in the exponential dependence on $k$. 

Instead, using the fact that $I_l$ is known to a very good approximation and further that information is approximately exponential in $k$, a good estimate for $I_k$ at large to intermediate $k$ is given by
\begin{equation*}
    I_k = 2^{-\lambda}I_{l} = 2^{-\lambda}\left( \log 2 - \frac{1}{2}  \right)\,.
\end{equation*}
This is the quantity quoted in \eqref{eq:mi_scaling_estimate_lln} in the main text.

\subsubsection{Moment scaling with $k$}
The following argument justifies \eqref{eq:moment_scaling_approximation} and is standard. We produce it here for completeness. Upon moving from $k$ to $k-1$ cooperants, the new conditional distribution is given by \eqref{marginalsk}. This means that $P_{k-1}$ is given by a convolution of $P_k$ with itself.
\begin{equation*}
    P_{k-1}(x) = \int_{\mathbb{R}^2} dy\, dz\, \delta\left(x - \frac{1}{2}(y + z)\right)P_k(y)P_k(z) = 2 \int_{\mathbb{R}} dy\,P_k(2x - y)P_k(y)
\end{equation*}
The characteristic function of a distribution over a continuous variable is its Fourier transform. Since $P_{k-1}$ is a convolution, its characteristic function is the product of characteristic function of $P_k$ with itself. A slightly more convenient object to work with is therefore the logarithm of the characteristic function:
\begin{equation*}
    \varphi_k(z) = \log \int_{\mathbb{R}} dx P_k(x) e^{izx}
\end{equation*}
The sum rule above gives a recursion relation for $\varphi_k$
\begin{equation*}
\varphi_{k-1}(z) = 2\varphi_{k}(z/2) + \log 2\,.    
\end{equation*}
Now, the cumulants of $P_k$ can be calculated from $\varphi_k$
\begin{equation*}
    c^{(a)}_k = \frac{1}{i^a}\frac{d^a}{dz^a}\varphi_k(z)\bigg|_{z=0}\,,
\end{equation*}
Meaning there are also recursion relations for the cumulants:
\begin{equation} \label{eq:cumulant_scaling_rule}
    c^{(a)}_{k-1} = 2^{1-a}c^{(a)}_{k}
\end{equation}
This leads directly to the central limit theorem, since when the second cumulant is rescaled to remain constant with respect to $n$, all higher-order cumulants scale to zero. Here, by using the fact that the $n^{\text{th}}$ moment $m^{(a)}_k$ can be expressed in terms of cumulants $c^{(b)}_{k}$ for $b\leq a$, we can see that the second-order cumulant dominates all of these expressions once $k$ is sufficiently small. For example, the fourth moment quickly scales like $(m^{(2)})^2$ as $k$ is decreased from $l$ because the second cumulant begins much larger than the fourth cumulant and remains dominant.
\begin{align*}
    c^{(2)}_{k} &= 2^{-\lambda} c^{(2)}_l = \frac{1}{12\cdot2^{\lambda}} \\
    c^{(4)}_{k} &= 2^{-3\lambda} c^{(4)}_l = \frac{-1}{120 \cdot 2^{3\lambda}}\\
    m^{(4)}_{k} &= c^{(4)}_k + 3 (c^{(2)}_k)^2 \approx 3 (c^{(2)}_k)^2 \sim 2^{-2\lambda}
\end{align*}
Hence, due to \eqref{eq:cumulant_scaling_rule}, $m^{(4)}_{k-1} \approx m^{(4)}_{k}/4$, which agrees with \eqref{eq:moment_scaling_approximation}.
\end{widetext}

\bibliographystyle{ieeetr}
\bibliography{refs}

\end{document}